\documentclass[twocolumn,showpacs,preprintnumbers,amsmath,amssymb,aps,prb]{revtex4}

\usepackage{graphicx}
\usepackage{dcolumn}
\usepackage{bm}
\usepackage{subfigure}
\usepackage{natbib}
\usepackage{multirow}

\begin{document}

\title{Ultrafast electrons dynamics reveal the high potential of InSe for hot carrier optoelectronics}

\author{Zhesheng Chen$^{1}$, Christine Giorgetti$^{1}$, Jelena Sjakste$^{1}$, Raphael Cabouat$^{1}$, Valerie Veniard$^{1}$, Zailan Zhang$^{2}$, Amina Taleb-Ibrahimi$^{3}$, Evangelos Papalazarou$^{4}$, Marino Marsi$^{4}$, Abhay Shukla$^{2}$, Jacques Peretti$^{5}$ and Luca Perfetti$^{1}$}

\affiliation{$^{1}$ Laboratoire des Solides Irradi\'{e}s, Ecole polytechnique, CNRS, CEA, Universit\'e Paris-Saclay, 91128 Palaiseau cedex, France}
\affiliation{$^{2}$ Institut de Min\'eralogie, de Physique des Mat\'eriaux et de Cosmochimie, Sorbonne Universit\'es - UPMC Univ Paris 06, CNRS-UMR7590, 4 Place Jussieu, Paris 75252, France}
\affiliation{$^{3}$ Soci\'{e}t\'{e} civile Synchrotron SOLEIL, L'Orme des Merisiers, Saint-Aubin - BP 48, 91192 GIF-sur-YVETTE, France}
\affiliation{$^{4}$ Laboratoire de Physique des Solides, CNRS, Universit\'{e} Paris-Sud, Universit\'{e} Paris-Saclay, 91405 Orsay, France}
\affiliation{$^{5}$ Laboratoire de Physique de la Matiere Condens\'ee, Ecole polytechnique, CNRS, CEA, Universit\'e Paris-Saclay, 91128 Palaiseau cedex, France}

\begin{abstract}
We monitor the dynamics of hot carriers in InSe by means of two photons photoelectron spectroscopy (2PPE). The electrons excited by photons of 3.12 eV experience a manifold relaxation. First, they thermalize to the electronic states degenerate with the $\bar M$ valley. Subsequently, the electronic cooling is dictated by Fr\"ohlich coupling with phonons of small momentum transfer. Ab-initio calculations predict cooling rates that are in good agreement with the observed dynamics. We argue that electrons accumulating in states degenerate with the $\bar M$ valley could travel through a multilayer flake of InSe with lateral size of 1 micrometer. The hot carriers pave a viable route to the realization of below-bandgap photodiodes and Gunn oscillators. Our results indicate that these technologies may find a natural implementation in future devices based on layered chalcogenides.
\end{abstract}


\maketitle

Van der Waals chalcogenides display a variety of different specifics that depend on their composition and number of layers. The weak mechanical binding of the atoms along the stacking direction facilitates the realization of heterostructures with different functionalities. Some recent achievements have been: the fine tuning of the band gap \cite{Heinz1,Mudd,Tran}, the control of valley polarization \cite{Heinz2} and the realization of devices with high mobility \cite{Kis,Li,Bandurin}. In this context, InSe is one of the building blocks with the highest potentials. 

The bulk crystals of InSe can be thinned down to a few layers and encapsulated in hBN \cite{Bandurin}. By these means, Bandurin and collaborators have fabricated transitors whose quality is high enough to observe Shubnikov - de Haas oscillations and quantum Hall effect \cite{Bandurin}. These results point out the two aspects making InSe particularly appealing. On one hand, the mobility of charge carriers rivals with the one measured in graphene \cite{Bolotin}.  On the other hand, the bulk band gap of 1.26 eV, is ideally suited for optoelectronic devices. Indeed, several groups have recently reported that InSe \cite{Lei,Tamalampudi} and InSe-graphene heterostructures \cite{Zhesheng} have excellent photoresponsivity in the visible spectral region. Such photodetectors could be patterned on a large scale over flexible supports \cite{Zheng}. Eventually, the application of larger bias can drive the photodetector in the avalanche regime \cite{Lei_Avalanche}. 

The conception of devices based on InSe would greatly profit from a precise knowledge of the transient state following photoexcitation. Some pioneering  experiments have investigated the coherent propagation and dephasing of the exciton-polariton \cite{Nusse,Dey}. These optical methods revealed a beating polarization induced by resonant pulses, but provided no insights on the energy relaxation of hot carriers. Here, we address this issue by mapping the dynamics of excited electrons in reciprocal space \cite{Ohtsubo,Hofmann}. Our two photon photoemission (2PPE) data show that photoexcitation above the $\bar M$ valley results in a high density of electrons with excess energy of $\cong0.6$ eV and cooling time of 2 ps. First principle calculations of the electron-phonon coupling discriminate between the distinct scattering mechanisms during the cooling process. The relevance of such hot carriers for the design of tunable photodetectors and Gunn oscillators is briefly discussed.

\begin{figure}
\includegraphics[width=0.9\columnwidth]{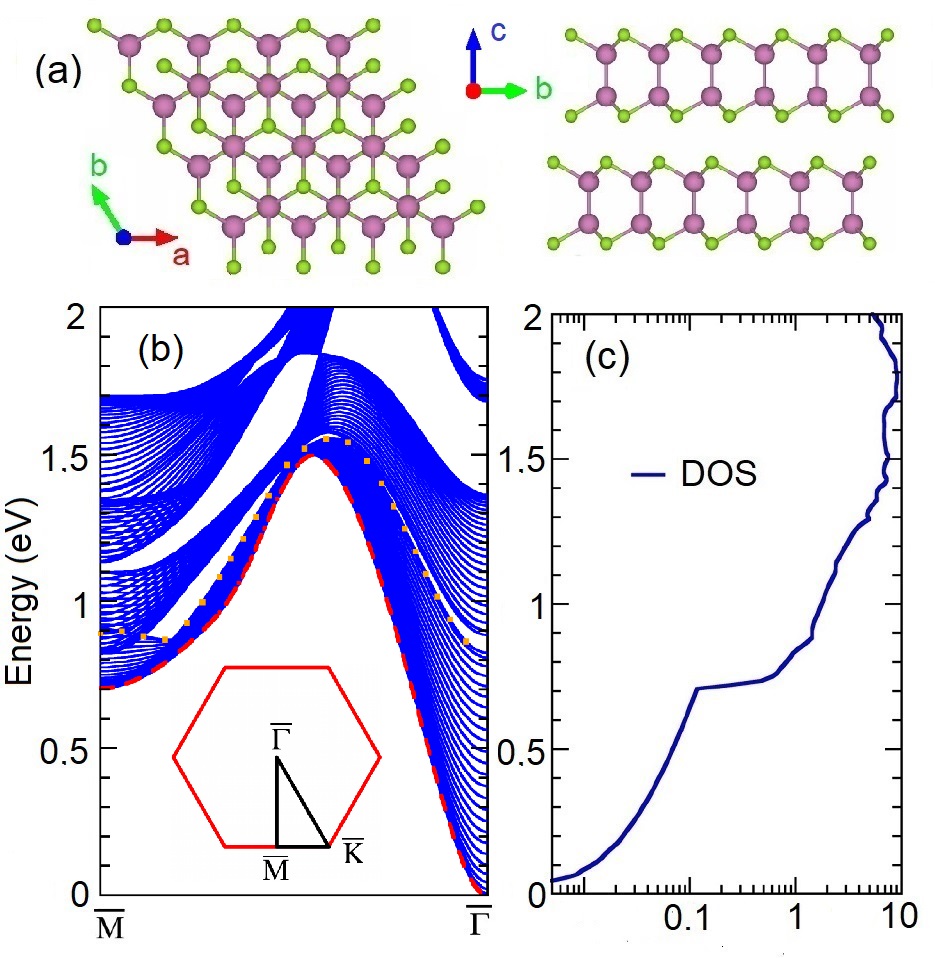}
\caption{a) Crystal structure of $\varepsilon$-InSe. b) Projected band structure along the high symmetry direction $\bar\Gamma - \bar M$. The projected states of the first conduction band span the area between the red dashed line and the orange dotted line. The inset display the surface Brillouin zone. c) Density of electronic states as a function of excess energy, in units of electron/cell/eV.}
\label{Fig1}
\end{figure}

Single crystals of $\varepsilon$-InSe have been grown using the Bridgmann method from a non-stoichiometric melt \cite{Chevy}. The crystallographic structure of $\varepsilon$-InSe is displayed inFig.\ref{Fig1}(a) (top and side views respectively left and right). Ultraviolet photoemission and photoluminescence spectra show that our bulk crystal is naturally n-doped and has direct band gap of 1.26 eV.  
We plot  in Fig. \ref{Fig1}(b) the calculated bulk band-structure projected on the surface plane, along the $\bar\Gamma-\bar M$ of the hexagonal Brillouin zone (BZ) (see inset). The dashed-red and dotted-orange lines point out the extension of the projected first conduction band. The correspondence between the bulk and surface BZ, a band structure along the complete $\bar\Gamma-\bar K-\bar M-\bar\Gamma$ path for a larger energy-range, as well as the technical details of the \textit{ab initio} calculations are given in the supplementary file. We plot in Fig. \ref{Fig1}(c) the projected density of electronic states calculated
by Density Functional Theory (DFT), using ABINIT code \cite{Gonze}, with lattice parameters from ref \cite{Magorrian}, (see Supplemental \cite{supplemental}). As show by Fig. \ref{Fig1}(c), the density of electronic states in the conduction band abruptly increases when the excess energy overcomes the bottom of the $\bar M$ valley.

Our photon source is a Ti:Sapphire laser system delivering 6 $\mu$J pulses with repetition rate of 250 kHz. Part of the fundamental beam ($\omega = 1.56$ eV) is converted to the second harmonic ($2\omega= 3.12$ eV) in a $\beta$-BBO crystal while the rest is employed to generate the third harmonic ($3\omega = 4.68$ eV) and fourth harmonic beam ($4\omega = 6.24$ eV) \cite{Faure}. We photoexcite the sample either by the $\omega$ or $2\omega$ pulses, generating carriers density of $8\times10^{17}$ cm$^{-3}$ and $7\times10^{18}$ cm$^{-3}$, respectively. The electrons emitted by the $3\omega$ or $4\omega$ pulse are discriminated in kinetic energy and angle with a resolution of 50 meV and $0.5$ degrees, respectively. Independently on the configuration of the experiment, the two beams generating the 2PPE signal display a cross correlation with FWHM $<0.16$ ps. All samples have been cleaved at the base pressure of $8\times 10^{-11}$ mbar and measured at temperature of 125 K.

\begin{figure}
\includegraphics[width=1\columnwidth]{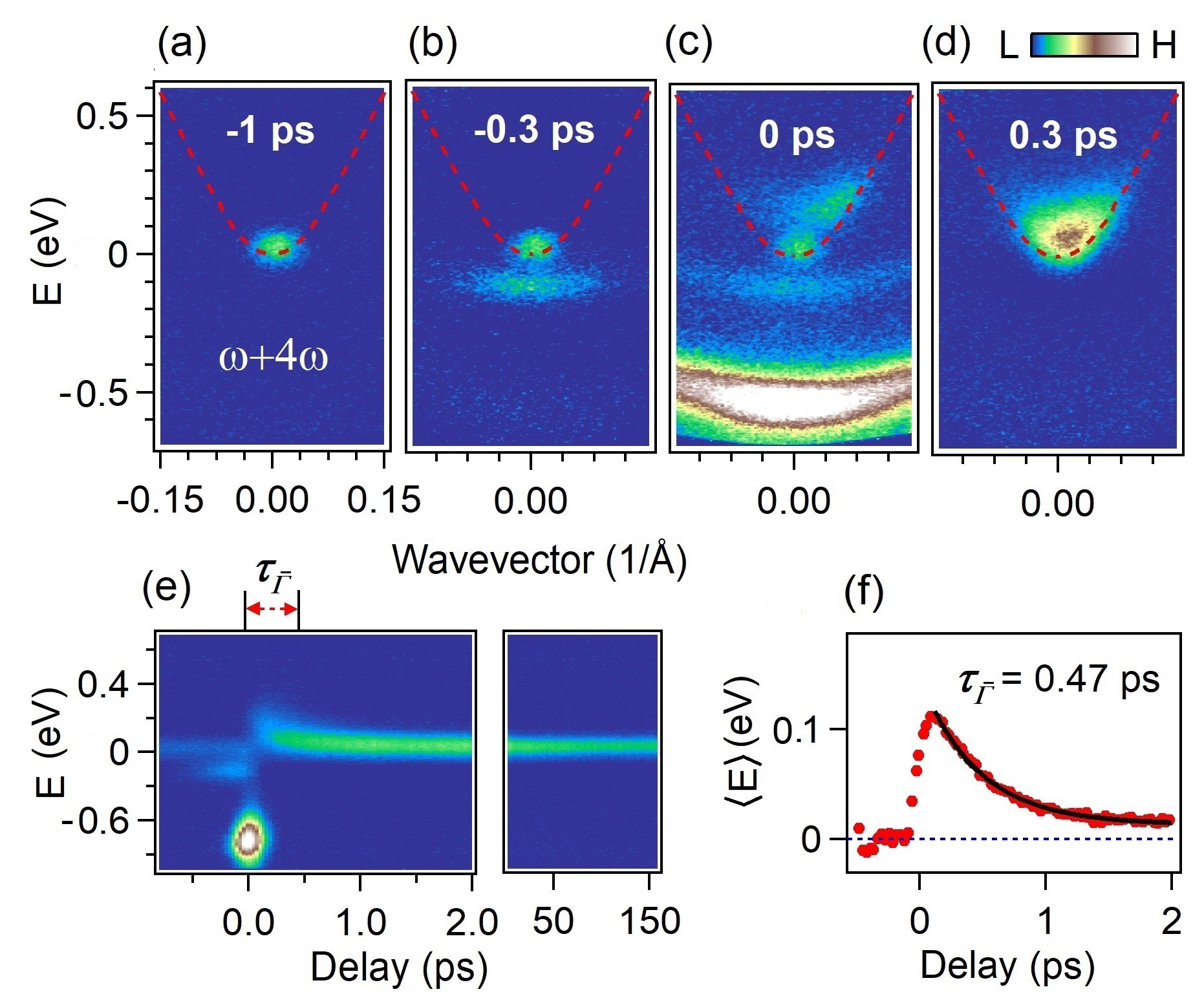}
\caption{The 2PPE data of this figure is generated by $\omega=1.56$ eV and $4\omega=6.24$ eV photons. (a-d) Photoelectron intensity maps acquired along the $\bar\Gamma-\bar M$ direction and plot as a function of excess energy for different delay times. The red dashed line is the conduction band dispersion. (e) Dynamics of photoelectron intensity integrated in the wavevector interval [-0.1, 0.1] \AA$^{-1}$. (f) Average excess energy of the electrons in the conduction band as a function of pump-probe delay. The solid line is an exponential fit with decay time $\tau_{\bar\Gamma}=0.47$ ps.}
\label{Fig2}
\end{figure}

Figure \ref{Fig2}(a-d) shows the photoelectron intensity maps acquired along the $\bar\Gamma-\bar M$ direction at different delay time $t$ between the $\omega$ pulse and $4\omega$ pulse. The image collected 1 ps before the arrival of the $\omega$ beam (Fig. \ref{Fig2}(a)) is representative of the system in equilibrium conditions. According to the nominal $n$-doping, we observe electrons in the conduction band minimun at excess energy $E=E_{\bar\Gamma}=0$ eV. 
Two image-potential-states below $E_{\bar\Gamma}$ are excited by the $4\omega$ beam and are detected, at negative $t$, by the $\omega$ pulse (Fig. \ref{Fig2}(b,c)). Due to the short lifetime, the lowest image-potential-state is clearly visible only in the intensity map of Fig. \ref{Fig2}(c). This map has been acquired at nominal value of zero delay and can be viewed as a snapshot of the primary excitations. Remark yet in Fig. \ref{Fig2}(c) that the $\omega$ pulse populates the conduction band up to $E=\omega-\Delta=0.25$ eV. The internal thermalization of such hot carriers proceeds faster than the cooling, probably because of an additional contribution of electron-electron scattering.

Figure \ref{Fig2}(e) shows the photoelectron intensity $I(E,t)$ integrated in the wavevector window [-0.1,0.1] \AA$^{-1}$ and plot as a function of pump probe delay. The average excess energy $\langle E \rangle$ of the electrons in the conduction band is obtained by evaluating the integral $\int E I(E,t) dE/\int I(E,t) dE$ in the interval $E\in [0,0.5]$ eV. As shown by Fig. \ref{Fig2}(f), the $\langle E \rangle$ has intial value $\langle E \rangle_0=0.12$ eV and it follows an exponential decay with time constant $\tau_{\bar\Gamma}=0.47 \pm 0.2$ ps. This timescale elucidates the main relaxation mechanism of hot carriers in the $\bar\Gamma$ valley. We recall that long-range Fr\"ohlich interaction diverges in a material with three dimensional periodicity. Therefore, the polar optical coupling becomes the dominant electron-phonon scattering channel for small momentum transfer \cite{Jelena,Tanimura2}.

Our calculations of vibrational properties and defomation potentials \cite{QuantumExpresso,Jelena1,Jelena2} (see Supplemental \cite{supplemental}) confirm the important role of polar interaction in the intravalley cooling. We estimate the energy dissipation rate $\lambda=d\langle E \rangle/dt$ between 0.21 and 0.24 eV/ps at 125 K . Due to the long-range nature of the Fr\"olich coupling, the phonon emission is weakly sensitive to the variation of electronic DOS. Therefore, $\lambda$ is almost independent on $\langle E \rangle$ as long as the excess energy is sufficiently larger than the phonon frequency. According to our theoretical estimate, hot carriers with an initial value $\langle E \rangle_0=0.12$ eV require a characteristic time $\langle E \rangle_0/\lambda=0.57$ ps to approach the asymptotic level. The reasonably good agreement between $\langle E \rangle_0/\lambda$ and the experimental $\tau_{\bar\Gamma}$ confirms that our simple analysis provides an accurate description of the observed cooling.

\begin{figure}
\includegraphics[width=1\columnwidth]{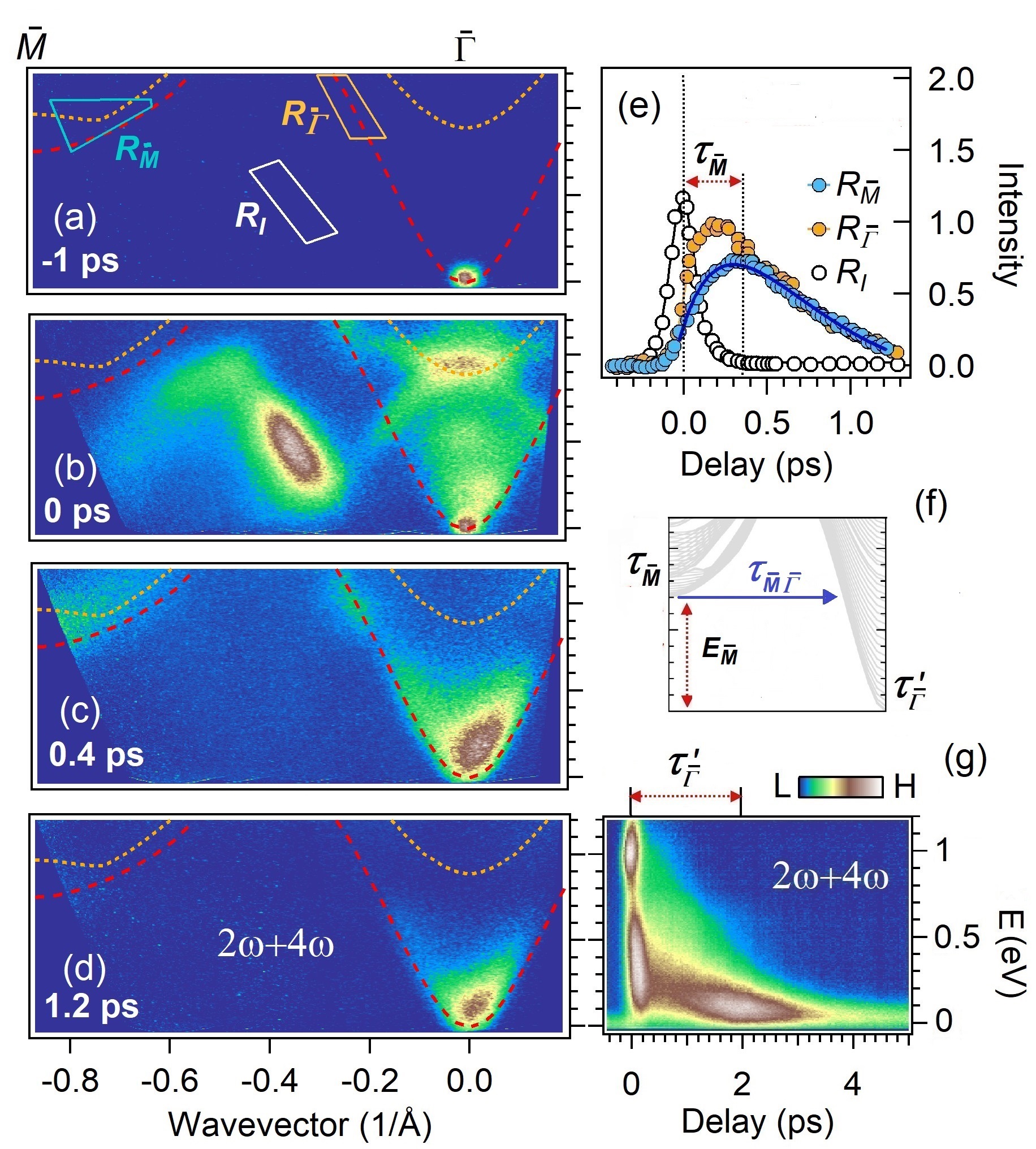}
\caption{The 2PPE data of this figure is generated by $2\omega= 3.12$ eV and $4\omega=6.24$ eV photons. (a-d) Photoelectron intensity maps acquired along the $\bar\Gamma-\bar M$ direction, and plot as function of excess energy for different delay times. The projected states of the conduction band span the area between the red dashed line and the orange dotted line. (e) Dynamics of photoelectron intensity in the $\bar M$ valley ($R_{\bar M}$), in the virtual intermediate state ($R_I$) and in the $\bar \Gamma$ valley ($R_{\bar \Gamma}$). The explicit integration areas of the photoemission signal are indicated in the upper panel of (a). (f) Sketch of the projected band structure with characteristic timescales of $\bar M$ valley, $\bar \Gamma$ valley thermalization and intervalley scattering. (g) Dynamics of photoelectron intensity integrated in the wavevector interval [-0.25,0.25] \AA$^{-1}$.}
\label{Fig3}
\end{figure}

Next, we investigate the photoexcited state generated by pumping and probing the sample with $2\omega$ and $4\omega$ pulses, respectively. We mark in Fig. \ref{Fig3}(a) the three indicative areas $R_{\bar M}$, $R_{\bar \Gamma}$ and $R_I$ on the 2PPE intensity map. As shown in Fig. \ref{Fig3}(b), the absorption of $2\omega$ photons results in a rich variety of spectral features. The projected states of the conduction band span between the red dashed line and the orange dotted line that are superimposed to the intensity maps. This guideline is useful to discriminate between different processes leading to photoemitted electrons. On one hand, the strong and dispersing signal that is located in the projected band gap is due to direct 2PPE emission via virtual states (very strong in $R_I$). Such photoelectrons may mimic the initial state dispersion and become intense when propagating in emitted waves that match the boundary conditions. On the other hand, real intermediate states of the $\bar\Gamma$ valley are excited by direct optical transitions from the $p_x$, $p_y$ orbitals of the valence band.
As in the case of GaAs, the efficient intervalley scattering can rapidly randomize the wavevector of electrons with high excess energy \cite{Tanimura1,Tanimura2}. This process sets in when $E$ overcomes the bottom of the $\bar M$ valley at $E_{\bar M}=0.7$ eV.  By comparing Fig. \ref{Fig3}(b) with Fig. \ref{Fig3}(c), we deduce that electrons have already reached $E_{\bar M}$ at $t=0.4$ ps (see evolution of the signal in $R_{\bar M}$). Furthermore, the intervalley scattering steadily refills the states of the $\bar \Gamma$ valley that are degenerate with $E_{\bar M}$ (signal in $R_{\bar \Gamma}$).

The dynamics of excited electrons can be tracked by focusing on selected regions in reciprocal space. We have integrated the 2PPE signal in the three indicative areas $R_{\bar M}$, $R_{\bar \Gamma}$ and $R_I$ of Fig. \ref{Fig3}(a). As shown by Fig. \ref{Fig3}(e), the photoemission assisted by virtual states ($R_I$) follows the cross correlation between the pump and probe pulse. This signal has been employed to estimate the duration of optical pulses and to precisely determine zero delay. Note that states at the bottom of the $ \bar M$ valley ($R_{\bar M}$) reach the highest occupation for $\tau_{\bar M}=0.36$ ps. Although these electrons loose the initial momentum within few tens of femtoseconds, their energy dissipation is restrained by the low quantum energy of the emitted phonons. The most relevant modes that are involved in the scattering process are acoustic branches close to the zone boundary and with typical energy of $\Omega\cong 14$ meV. Our calculations indicate that electrons with $E-E_{\bar M}\cong 0.5$ eV scatter with such phonons on an average timescale of  8 fs at 125 K. The energy dissipation time at high excess energies is mostly determined by short-wavelength phonon scattering, which depends strongly on the excess energy. According to our theoretical estimates the electrons that are at 0.5 eV above $E_{\bar M}$ require a characteristic time constant of 0.2 ps to reach the bottom of the $\bar M$ valley (see Supplemental \cite{supplemental}).

Figure \ref{Fig3}(e) shows also that $R_{\bar M}$ electrons in the $\bar M$ valley and $R_{\bar \Gamma}$ electrons in the $\bar \Gamma$ valley display an identical dynamics for $t>\tau_{\bar M}$. These nearly-degenerate states are maintained in detailed balance conditions by phonons with wavevector near to the $\bar M$ point. The transition time calculated at 125 K is $\tau_{\bar M \rightarrow \bar \Gamma}=180 $ fs and it turns out to be short enough to confirm our conjecture. Eventually, the large $E_{\bar M}=0.7$ eV may suggest that 180 fs (or 80 fs at 300 K) is surprisingly long. This result does not derive from unconventional matrix elements of electron-phonon coupling but rather from the small DOS of the $\bar \Gamma$ valley at $E_{\bar M}$ (See Fig. \ref{Fig1}(c)). We find instructive the comparison of InSe with other chalcogenides: electrons that are resonantly excited at 300 K in the $\bar M$ valley of bulk MoS$_2$ \cite{Wallauer} or WSe$_2$ \cite{Bertoni} require $\tau_{\bar M \rightarrow \bar \Sigma}< 80$ fs to reach the $\bar \Sigma$ valley and the energy separation between $\bar M$ and $\bar \Sigma$ valley is only $E_{\bar M}-E_{\bar \Sigma}=0.2$ eV. In the case of InSe, the excess energy is particularly large and therefore the carriers can be sustained above the conduction band minimum for long time. Moreover, the abrupt increase of the DOS for $E>E_{\bar M}=0.7$ eV, can boost the injection of electrons in the $\bar M$ valley. Once these electrons fall below $E_{\bar M}$, the intervalley scattering is no longer active whereas the cooling proceeds mainly via Fr\"ooling coupling to phonons with small momentum transfer \cite{Tanimura1,Tanimura2}. Figure \ref{Fig3}(g) shows the photoelectron signal integrated around the $\bar \Gamma$ valley in a wavevector window of [-0.25, 0.25] \AA$^{-1}$. The maximal carrier density at the bottom of the conduction band is achieved $\tau_{\bar \Gamma'}=2$ ps after photoexcitation.

\begin{figure}
\includegraphics[width=0.7\columnwidth]{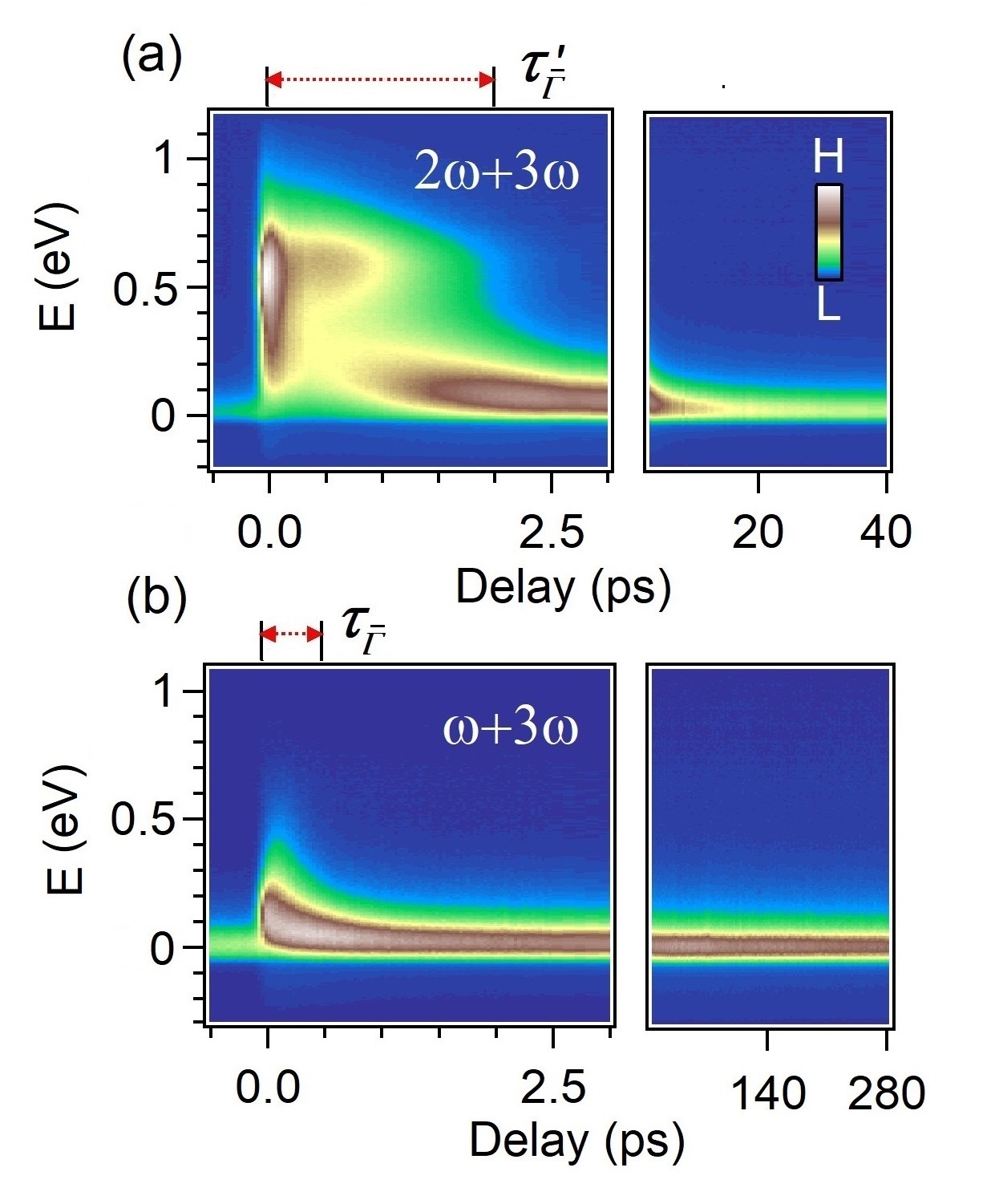}
\caption{The 2PPE data of this figure have been collected by probing photoelectrons with $3\omega=4.68$ eV photons. Dynamics of electrons excited in the $\bar \Gamma$ valley by photons of (a) $2\omega=3.12$ eV and (b) $\omega=1.56$ eV. Long living hot carriers take place only upon photoexcitation with $2\omega$ pulses.}
\label{Fig4}
\end{figure}

Figure \ref{Fig4}(a) and Fig. \ref{Fig4}(b) display the 2PPE signal probed in the $\bar \Gamma$ valley by 3$\omega$ photons and excited with 2$\omega$ and $\omega$, respectively. Being independent on the DOS, the Fr\"olich dissipation rate is the same in both cases. However, the electrons hailing from the $\bar M$ valley gather in the $\bar \Gamma$ valley near to the calculated $E_{\bar M}$. As a consequence, the energy dissipation takes much longer in Fig. \ref{Fig4}(a) than in Fig. \ref{Fig4}(b). By assuming that intervalley scattering generate an average kinetic energy $\langle E \rangle\cong 0.4-0.6$ eV, the resulting cooling time $\langle E \rangle/\lambda=1.6-2.5$ ps is consistent with the observed $\tau_{\bar \Gamma'}$.

Next, we estimate the minimal distance $d$ at which two electrodes should be placed in order to extract the hot carriers. The effective mobility $\mu=600 $ cm$^2$V$^{-1}$s$^{-1}$ is given by the model for linear dispersion \cite{Avouris} at 0.7 eV excess energy and 125 K (see Supplemental \cite{supplemental}). Due to the high excess energy of hot electrons, the nominal $\mu$ is lower than the mobility measured in InSe transistors \cite{Bandurin}. Nonetheless a driven drift can still attain high values. Upon an applied voltage of $V=10$ Volts, the hot  electrons reach the electrodes if $d=\sqrt{V\mu \tau'_{\bar \Gamma}} \cong 1 \mu m$.

The relatively large distance that can be covered by hot carriers suggests original concepts for optoelectronic applications. Recently Lao \emph{et al.} developed a long-wavelength photodetection principle working far from equilibrium conditions \cite{Lao}. Hot carriers injected into $p$-doped GaAs structure interact with cold carriers and promote them to states of high excess energy. The tunneling of such holes through Al$_x$Ga$_{1–x}$As barriers enables a long-wavelength infrared response that can be tuned via the applied voltage. By direct analogy, the hot electrons in the $\bar M$ valley of InSe could extend such promising approach to chalcogenides heterostructures.

A second possible application of hot carriers generation are Gunn diodes. In III/V semiconductors, a microwave signal of high frequency appears if the applied bias field exceeds a threshold value \cite{Butcher}. This effect arises when the high electrical field promotes electrons from the conduction band minimum to an upper valley \cite{Butcher,McCumber}. As a consequence the average mobility of electrons decreases, while the device enters in a regime of negative differential resistance. The heavy mass of $\bar M$ valley should favor this process also in InSe. If confirmed, our conjecture may lead to the development of microwave oscillators and bistable switches based on chalcogeniges materials.

In conclusion, we have shown that InSe has excellent potential for hot carrier optoelectronics. The high energy density of electronic states above the bottom of $\bar M$ valley favors the photoassisted or electrical injection of electrons with high excess energy. On the other hand, electrons that reached the $\bar \Gamma$ valley can dissipate energy only via Fr\"ohlich coupling to optical phonons. These hot carriers have typical lifetime of 2 picoseconds and could be extracted from a device of $\sim 1 $ micron. We could successfully model the observed dynamics by first principle calculations of the electron-phonon scattering. Our results will be of guidance for the physics of electron-phonon coupling in other layered chalcogenides. Moreover, we suggest appealing concepts, such as below band-gap photodetection or negative differential resistance, that could be shortly integrated in to the family of van der Waals semiconductors.

We thanks DIM-Oximore and the Ecole Polytechnique for funding under the project 'ECOGAN'.
This work was supported by LABEX PALM (ANR-10-LABX-0039-PALM, Project 
Femtonic) and by the EU/FP7 under the contract Go Fast (Grant No. 280555). Computer time has been granted by GENCI (project 2210) and by Ecole 
Polytechnique through the LLR-LSI project.

\end{document}